\documentstyle[twocolumn,aps]{revtex}

\begin{document}
\draft
\title{Magneto-exciton in single and coupled type II quantum dots}
\author{K. L. Janssens\cite{karenmail}, B. Partoens\cite{bartmail} and F. M. Peeters 
\cite{peetersmail}}
\address{Departement Natuurkunde, Universiteit Antwerpen (UIA), Universiteitsplein1,\\
B-2610 Antwerpen, Belgium}
\date{\today}
\maketitle

\begin{abstract}
We studied the exciton energy in a type II quantum disk as a function of the
magnetic field, disk radius $R$ and height $d$. We found angular momentum
transitions for dots with $d>>2R.$ Application of an electric field
perpendicular to the disk showed a non-linear Stark shift. In the case of
three vertically coupled dots angular momentum transitions were found for
small interdot distances which disappeared with increasing interdot distance.
\end{abstract}


\section{Introduction}

Self-assembled quantum dots, as realized by the Stranski-Krastanow growth
mode, have attracted considerable interest during the last decade. Most of
this interest was dedicated to the study of type I quantum dots \cite
{wang,bayer,stier,karen1}, e.g. $InAs/GaAs,$ where both electron and hole
are confined in the dot. Also very interesting, though less studied, are the
type II quantum dots \cite{hayne,sugisaki,karen2,lelong}, where the quantum
dot forms an antidot for one of the carriers, e.g. for the holes in
typically the $InP/GaInP$ system or the electrons in e.g. $GaSb/GaAs.$

In the first part of this work, we focus our attention on the properties of
an exciton in a single $InP/GaInP$ quantum dot, where the hole is located in
the barrier material. Our quantum dot is modeled as a quantum disk, with
radius $R$ and thickness $d$ and strain effects are neglected. As a
confinement potential, we take a hard wall of finite height. Furthermore, we
apply an external magnetic field in the growth direction. To solve the
problem, we use a Hartree-Fock (HF) mesh calculation, which allows very
flexible solutions.

The second part of this work was dedicated to the study of an exciton in
three vertically coupled identical dots with an interdot distance $d_{d}$.

\section{Theoretical model}

We obtained the exciton energy and wavefunction by solving the HF single
particle equations within the effective mass approximation (with $m_{e}$ and 
$m_{h}$ the effective electron and hole masses, respectively, $r_{e,h}=\sqrt{%
x_{e,h}^{2}+y_{e,h}^{2}},$ $\omega _{c,e}=eB/m_{e}$ and $\omega
_{c,h}=eB/m_{h}$). The HF single particle equations can be written as 
\begin{eqnarray}
\left[ -\frac{\hbar ^{2}}{2m_{e}}\frac{1}{r_{e}}\frac{\partial }{\partial
r_{e}}\left( r_{e}\frac{\partial }{\partial r_{e}}\right) -\frac{\hbar ^{2}}{%
2m_{e}}\frac{\partial }{\partial z_{e}}+\frac{\hbar ^{2}}{2m_{e}}\frac{%
l_{e}^{2}}{r_{e}^{2}}+\frac{l_{e}}{2}\hbar \omega _{c,e}+\frac{1}{8}%
m_{e}\omega _{c,e}^{2}r_{e}^{2}\right. &&  \nonumber \\
\left. +V_{e}(r_{e},z_{e})-\frac{e^{2}}{4\pi \epsilon }\int \frac{\rho
_{h}(r^{\prime },z^{\prime })}{|{\bf r}-{\bf r}^{\prime }|}d{\bf r}^{\prime }%
\right] \psi _{e}(r_{e},z_{e})=\epsilon _{e}\psi _{e}(r_{e},z_{e}), && 
\eqnum{1a} \\
\left[ -\frac{\hbar ^{2}}{2m_{h}}\frac{1}{r_{h}}\frac{\partial }{\partial
r_{h}}\left( r_{h}\frac{\partial }{\partial r_{h}}\right) -\frac{\hbar ^{2}}{%
2m_{h}}\frac{\partial }{\partial z_{h}}+\frac{\hbar ^{2}}{2m_{h}}\frac{%
l_{h}^{2}}{r_{h}^{2}}-\frac{l_{h}}{2}\hbar \omega _{c,h}+\frac{1}{8}%
m_{h}\omega _{c,h}^{2}r_{h}^{2}\right. &&  \nonumber \\
\left. +V_{h}(r_{h},z_{h})-\frac{e^{2}}{4\pi \epsilon }\int \frac{\rho
_{e}(r^{\prime },z^{\prime })}{|{\bf r}-{\bf r}^{\prime }|}d{\bf r}^{\prime }%
\right] \psi _{h}(r_{h},z_{h})=\epsilon _{h}\psi _{h}(r_{h},z_{h}), && 
\eqnum{1b}
\end{eqnarray}
with $\rho _{e}(r^{\prime },z^{\prime })$ and $\rho _{h}(r^{\prime
},z^{\prime })$ respectively the electron and hole densities. As confinement
potentials, we take hard walls of finite height, i.e. 
\begin{equation}
V_{e(h)}(r_{e(h)},z_{e(h)})=\left\{ 
\begin{array}{c}
V_{e(h)},\;\;r_{e(h)}>R\text{ and\ }\left| z_{e,h}\right| >d/2, \\ 
0,\;\;\mbox{otherwise},
\end{array}
\right.  \eqnum{2}
\end{equation}
with $R$ the radius of the disk and $d$ its thickness.

The equations have to be solved self-consistently, which is done
iteratively. We start with the free electron solution because in the absence
of any Coulomb interaction only the free electron is confined. The Hartree
integrals are integrated numerically 
\begin{equation}
\int \frac{\rho (r^{\prime },z^{\prime })}{|{\bf r}-{\bf r}^{\prime }|}d{\bf %
r}^{\prime }=4\int dz^{\prime }\int \frac{\rho (r^{\prime },z^{\prime
})r^{\prime }}{\sqrt{\left( r+r^{\prime }\right) ^{2}+\left( z-z^{\prime
}\right) ^{2}}}{\cal K}\left( \frac{4rr^{\prime }}{(r+r^{\prime
})^{2}+\left( z-z^{\prime }\right) ^{2}}\right) dr^{\prime },  \eqnum{3}
\end{equation}
where ${\cal K}(x)$ is the complete elliptic integral of the first kind.
More details about the calculation and numerical implementation of this
integral is given in \cite{karen2}.

After convergence of the iteration procedure, the total energy is obtained
as follows 
\begin{equation}
E_{\mbox{exciton}}=\epsilon _{e}+\epsilon _{h}+\frac{e^{2}}{4\pi \epsilon }%
\int \int \frac{\rho _{e}(r,z)\rho _{h}(r^{\prime },z^{\prime })}{|{\bf r}-%
{\bf r}^{\prime }|}d{\bf r}d{\bf r}^{\prime },  \eqnum{4}
\end{equation}
where $\epsilon _{e(h)}$ is the electron (hole) single particle energy and $%
\rho _{e(h)}$ is the electron (hole) density.

\section{Results}

First we considered the exciton in a single quantum disk. We took the
following material parameters: $m_{e}=0.077m_{0},$ $m_{h}=0.60m_{0},$ $%
V_{e}=250meV$, $V_{h}=-50meV$ and $\epsilon =12.61.$ The influence of the
disk parameters (i.e. radius and height) were studied in the absence of a
magnetic field which resulted in the phase diagram shown in Fig.~1. The
solid curve corresponds to the probability to find 50\% of the hole
wavefunction at the radial boundary of the disk. Here the probability was
calculated as 
\[
P_{side}=2\pi \int_{-\infty }^{\infty }dz_{h}\int_{R}^{\infty }r_{h}\left|
\Psi _{h}(r_{h},z_{h})\right| ^{2}dr_{h}. 
\]
One can distinguish between two regimes: a) the disk-like regime $\left(
d<<2R\right) $ where less than 50\% of the hole wavefunction is situated at
the radial boundary, and b) the pillar-like regime $\left( d>>2R\right) $
for the case of $P_{side}>50\%.$ In order to give a more visual picture, we
made contourplots of the hole wavefunction for the two regimes, shown as
insets to Fig.~1 for the cases of $R=4nm,$ $d=12nm$ and $R=12nm,$ $d=4nm.$
The two regimes are a consequence of the fact that the hole tends to sit as
closely to the electron as possible. Therefore, when $d<<2R,$ the hole will
prefer to sit above and below the dot, whereas for $d>>2R$ the hole will be
located at the radial boundary. Because the volume available for the hole
above and below the dot is smaller than at the radial boundary we found that
the 50\% division line is not given by $d\simeq 2R$ but could be
approximated by $d\simeq -8+2R$ in the investigated $\left( d,R\right) $
range as presented in Fig.~1.

An exciton will behave differently when a magnetic field is applied along
the $z$-direction depending on the position of the quantum dot in the phase
diagram. Fig.~2 shows the exciton energy as function of the magnetic field
for a disk-like system (Fig.~2(a), $R=12nm,$ $d=4nm$) and for a pillar-like
system (Fig.~2(b), $R=4nm,$ $d=12nm$). We find that in the first case the
angular momentum of the hole, $l_{h},$ remains zero over the whole $B$%
-regime, whereas for the latter case we find angular momentum transitions
with increasing magnetic field. These follow from the fact that, when the
hole is sitting at the radial boundary, the magnetic field pushes the hole
closer to the disk boundary, making it energetically more favourable to jump
to a higher $l_{h}$ state.

In a next step, we consider the influence of an electric field, applied
along the $z$-direction. Application of such an electric field leads to a
polarisation of the system. Following parameters were used: $R=10nm,$ $%
d=2nm, $ $V_{e}=250meV$ and $V_{h}=-50meV.$\ Fig.~3 depicts the dependence
of the exciton energy on the applied electric field. We observe a shift
towards lower energies with increasing $F,$ which is the so called Stark
shift. The inset of Fig.~3 shows the behaviour of the hole wavefunctionwhich
moves towards the bottom of the disk with increasing electric field. The
Stark shift is almost entirely due to the shift of the hole. Note that the
calculation is limited to very small values of the electric field. Since the
hole is only confined to the dot by the Coulomb attraction to the electron,
the system becomes fastly unbound at higher fields.

Finally, the system of three vertically coupled dots was studied under the
influence of a magnetic field. This system is especially interesting as it
allows to construct a pillar-like structure. We now have an extra parameter
to vary, namely the interdot distance $d_{d}.$ When $d_{d}$ is very small,
there will be not enough space for the holes to sit between the quantum
dots. Therefore, when the total stack height (i.e. three times the dot
thickness plus two times the interdot distance) is larger than the disk
diameter, the hole will prefer to sit at the radial boundary, exactly as in
the case of the pillar-like system for a single dot. For our numerical work
we used the parameters: $R=8nm,$ $d=3nm,$ $V_{e}=250meV$ and $V_{h}=-30meV.$%
\ Fig.~4 shows the exciton energy for $d_{d}=3nm$ as function of the
magnetic field, and we see again the appearance of angular momentum
transitions. The insets of Fig.~4 show contourplots for the hole
wavefunction at respectively $B=0T,$ where $l_{h}=0,$ and $B=50T,$ where $%
l_{h}=4$ is the groundstate. At $B=0T,$ part of the hole wavefunction is
still between the dots. By applying a strong magnetic field, a large part of
the hole wavefunction is pushed between the dots. However, there is not
enough space, and therefore it is more favourable to jump to a higher $l_{h}$
state. When we increase now the interdot distance $d_{d}$ to $5.5nm,$ we
give the hole more space between the dots and in this way create a
transition to a regime without angular momentum transitions (like the
disk-like regime). This is depicted in Fig.~5, where the insets show the
contourplots of the hole wavefunction at respectively $B=0T$ and $B=50T.$ We
find no angular momentum transition, and we see that at $B=50T$ the hole is
entirely situated between the dots.

\section{Conclusions}

We studied the exciton energy in a single type II quantum disk as a function
of the magnetic field. We made a distinction between a disk-like regime,
where the hole angular momentum remains zero, and a pillar-like regime,
where angular momentum transitions appear with increasing magnetic field.
Applying an electric field results in a Stark shift towards lower energies
and a strong polarisation of the hole wavefunction. Finally we studied the
case of three vertically coupled dots. For very small interdot distances,
this systems resembles the pillar-like regime, with the occurrence of
angular momentum transitions. When increasing the interdot distance, we
found no longer $l_{h}$ transitions, because of the possibility for the hole
to sit between the dots.

\section{Acknowledgements}

This work is supported by the Flemish Science Foundation (FWO-Vl), The
Inter-University Attraction Poles research program (IUAP-IV) and the
University of Antwerp (GOA and VIS). K. L. J. is supported by the
``Instituut voor de aanmoediging van Innovatie door Wetenschap en
Technologie in Vlaanderen'' (IWT-Vl) and B. P. is a post-doctoral researcher
with the FWO-Vl. Discussions with M. Tadic, M. Hayne and V. Moshchalkov are
gratefully acknowledged.

\bigskip

\begin{figure}[tbp]
\caption{Phase diagram of the probability for the hole wavefunction to sit
at the radial border of the disk, as a function of both $R$ and $d.$ The
insets show the contourplots of the hole wavefunction, at respectively $%
R=4nm,$ $d=12nm$ and $R=12nm,$ $d=4nm$. The dashed lines indicate the
boundary of the disk.}
\label{Fig1}
\end{figure}

\begin{figure}[tbp]
\caption{The exciton energy as a function of the magnetic field, for (a) a
disk-like structure and (b) a pillar-like structure.}
\end{figure}

\begin{figure}[tbp]
\caption{The exciton energy as a function of a vertical electric field. The
inset shows the hole wavefunction for a field of $F=1kV/cm.$}
\end{figure}

\begin{figure}[tbp]
\caption{The exciton energy as a function of the magnetic field for three
coupled dots, with $d_{d}=3nm.$ The insets are contourplots of the hole
wavefunction at respectively $B=0T$ and $B=50T.$}
\end{figure}

\begin{figure}[tbp]
\caption{The same as Fig.~4, but now for $d_{d}=5.5nm.$}
\end{figure}


\bigskip

\begin{references}
\bibitem[*]{karenmail}  corresponding author, electronic mail:
karenj@uia.ua.ac.be

\bibitem[%
\circ%
%
]{bartmail}  Electronic mail: bpartoen@uia.ua.ac.be

\bibitem[%
\dagger%
%
]{peetersmail}  Electronic mail: peeters@uia.ua.ac.be

\bibitem{wang}  P.D. Wang, J.L. Merz, S. Fafard, R. Leon, D. Leonard, G.
Medeiros-Ribeiro, M. Oestreich, P.M. Petroff, K. Uchida, N. Miura, H.
Akiyama, and H. Sakaki, Phys. Rev. B {\bf 53}, 16458 (1996).

\bibitem{bayer}  M. Bayer, A. Schmidt, A Forchel, F. Faller, T.L. Reinecke,
P.A. Knipp, A.A. Dremin, and V.D. Kulakovskii, Phys. Rev. Lett. {\bf 74},
3439 (1995).

\bibitem{stier}  O. Stier, M. Grundmann, and D. Bimberg, Phys. Rev. B {\bf 59%
}, 5688 (1999).

\bibitem{karen1}  K.L. Janssens, F.M. Peeters, and V.A. Schweigert, Phys.
Rev. B {\bf 63}, 205311 (2001).

\bibitem{hayne}  M. Hayne, R. Provoost, M.K. Zundel, Y.M. Manz, K. Eberl,
and V.V. Moshchalkov, Phys. Rev. B {\bf 62}, 10324 (2000).

\bibitem{sugisaki}  M. Sugisaki, H.-W. Ren, K. Nishi, S. Sugou, T. Okuno,
and Y. Masumoto, Physica B {\bf 256-258}, 169 (1998).

\bibitem{karen2}  K.L. Janssens, B. Partoens, and F.M. Peeters, accepted for
publication in Phys. Rev. B.

\bibitem{lelong}  Ph. Lelong, K. Suzuki, G. Bastard, H. Sakaki, and Y.
Arakawa, Physica E {\bf 7}, 393 (2000).
\end{references}
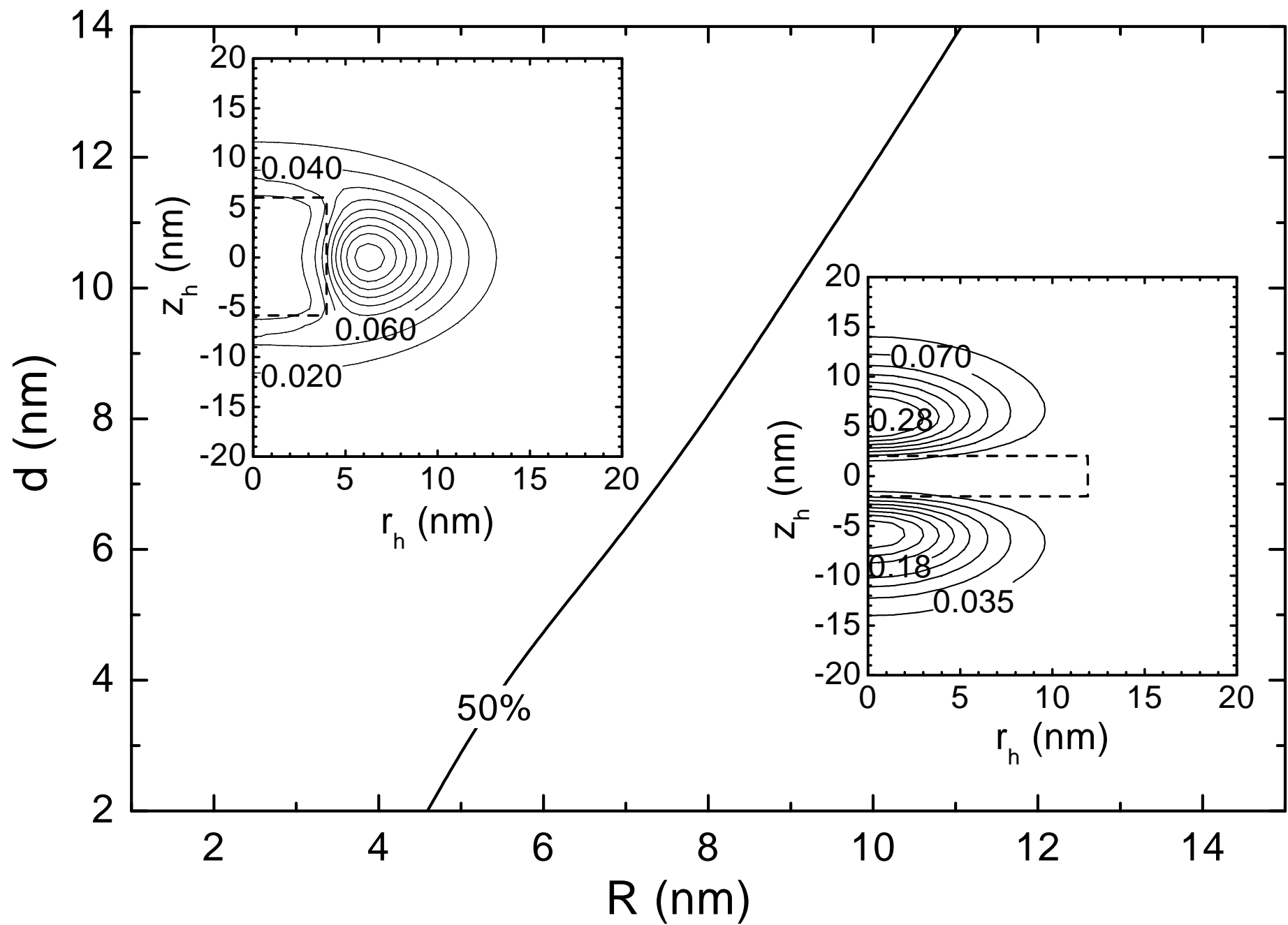
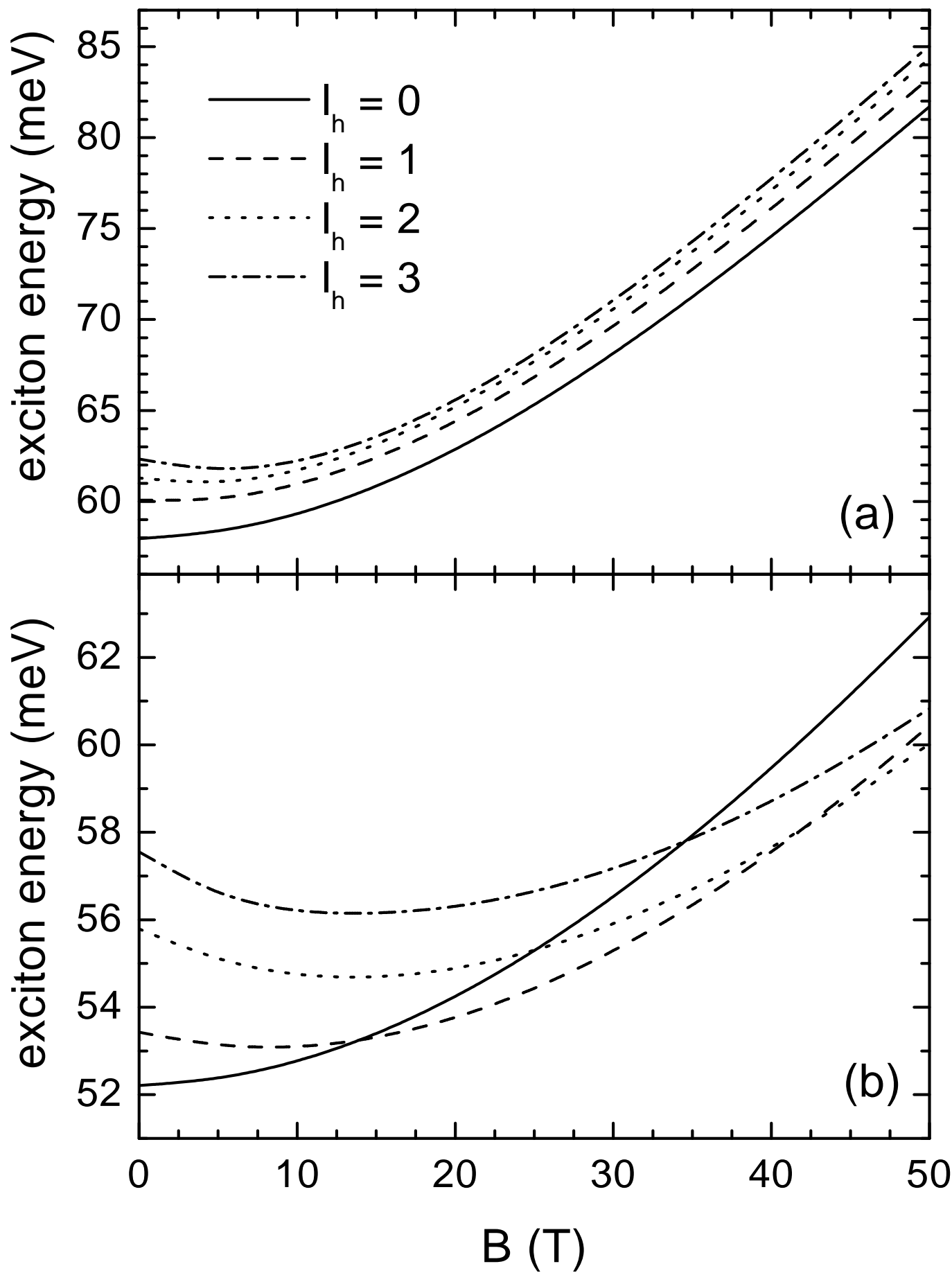
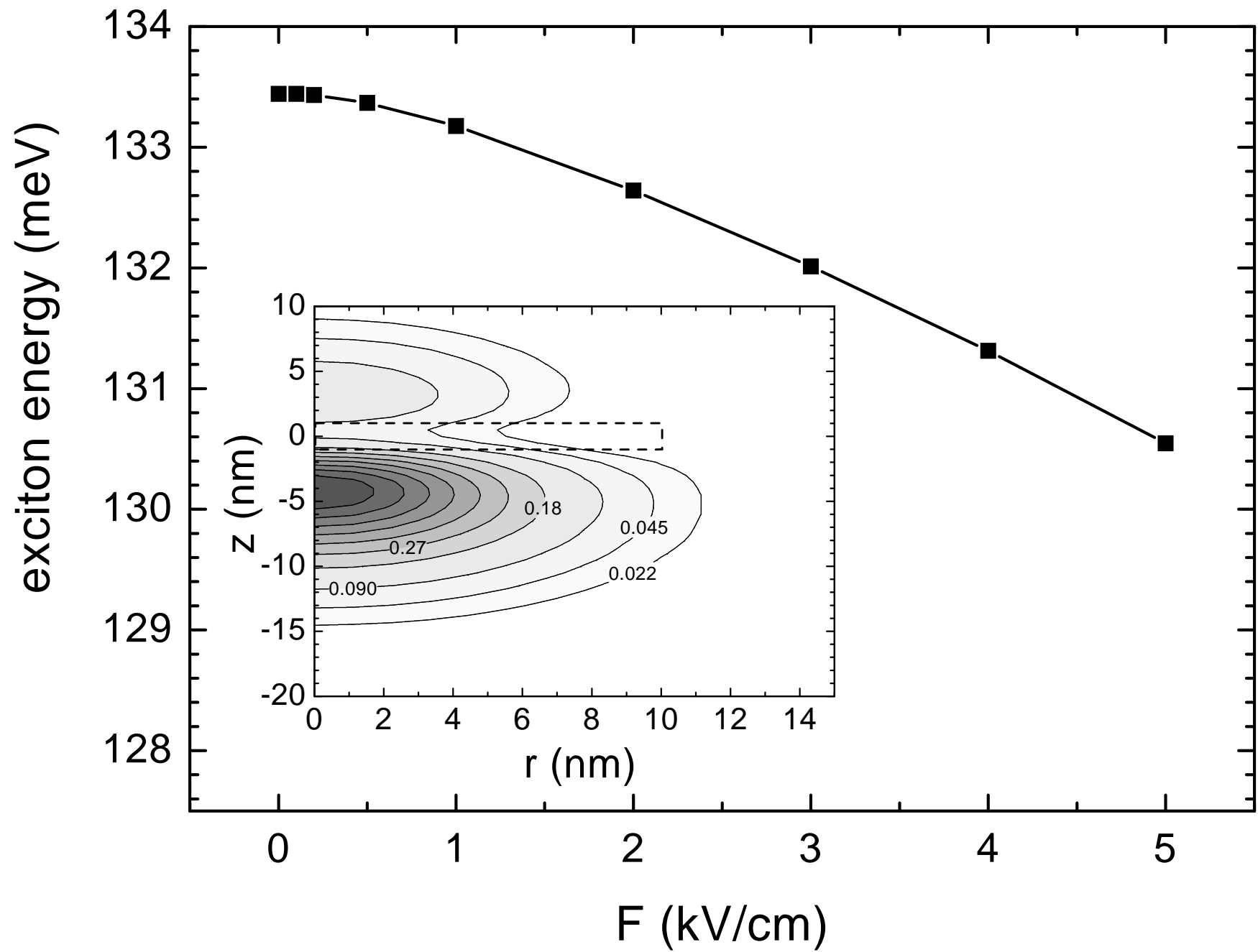
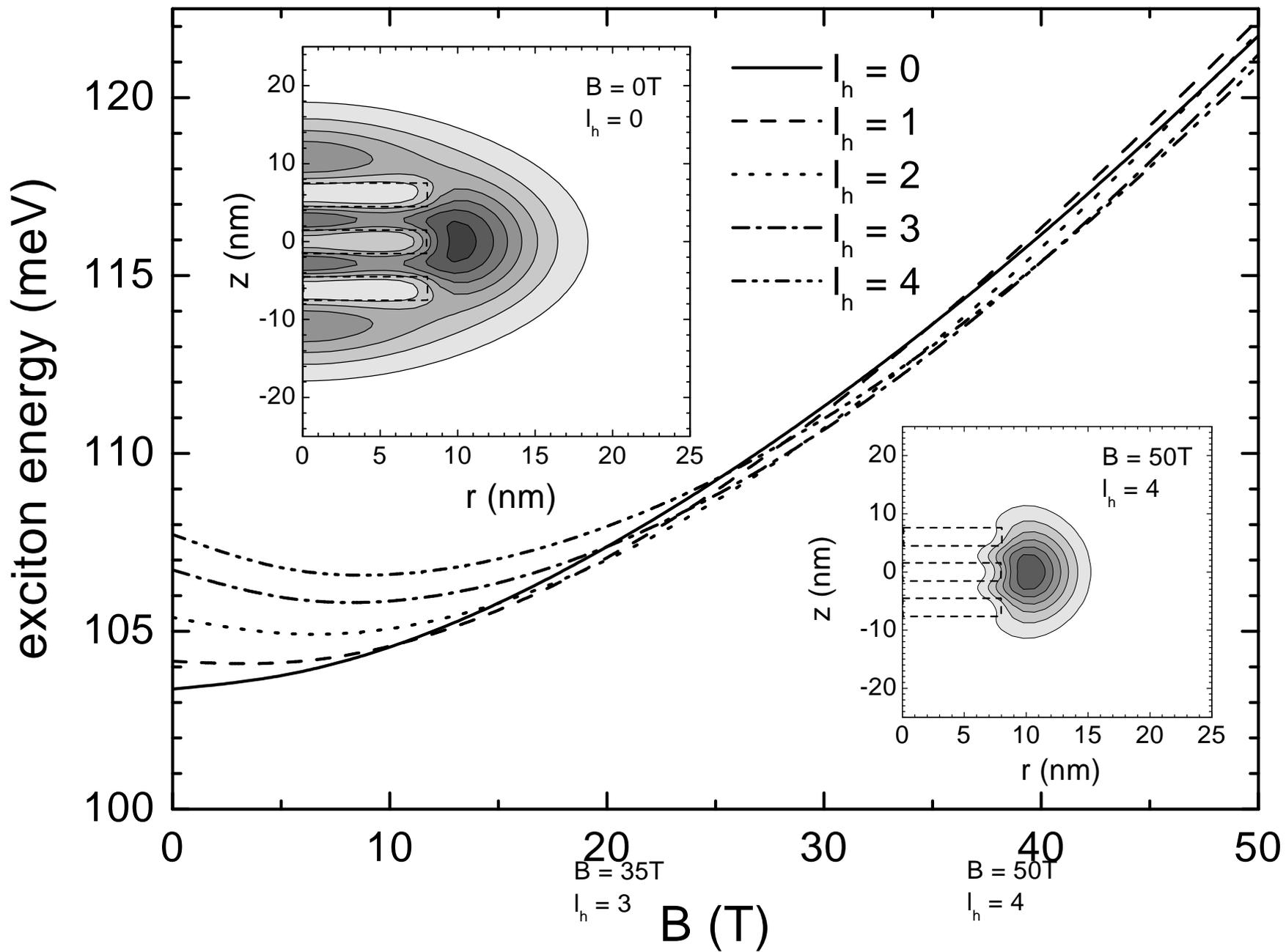
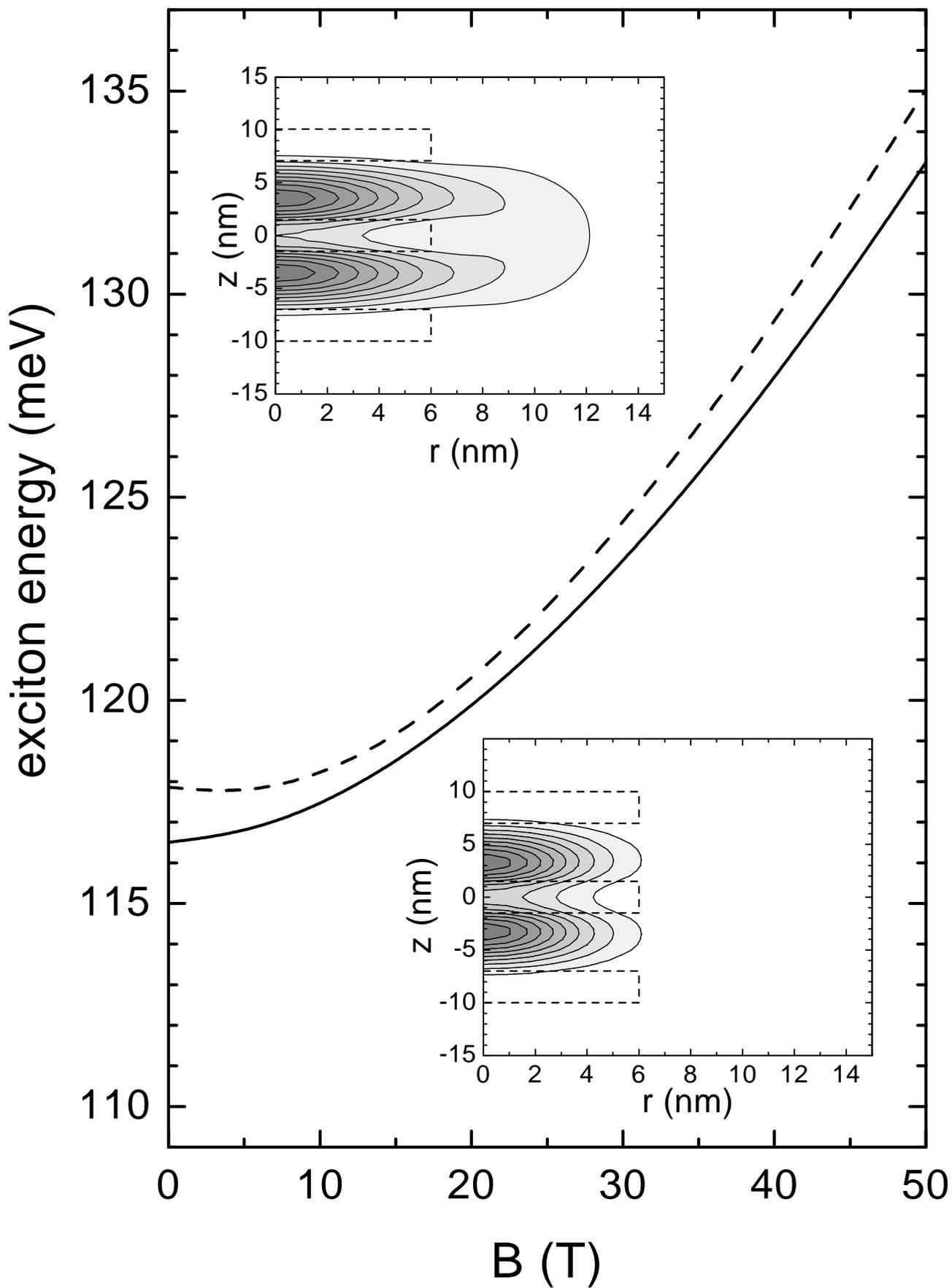
\end{document}